\documentclass[prb,preprint,showpacs,preprintnumbers,amsmath,amssymb]{revtex4}
\usepackage{graphicx}% Include figure files
\usepackage{dcolumn}% Align table columns on decimal point
\usepackage{bm}% bold math

\begin{document}

%\preprint{APS/123-QED}

\title{Magnetic ordering of Mn sublattice, dense Kondo lattice behavior of Ce in (RPd$_3$)$_8$Mn (R = La, Ce)}% Force line breaks with \\

\author{Surjeet Singh}
\email{surjeets@tifr.res.in}
%\altaffiliation{CMP\&MS, TIFR, Homi Bhabha Road, Mumbai 400 005, India} \\
\author{S K Dhar}
\email{sudesh@tifr.res.in}
\affiliation{CMP\&MS, TIFR, Homi Bhabha Road, Mumbai 400 005, India} % \\

\date{\today}

\begin{abstract}
We have synthesized two new interstitial compounds (RPd$_3)_8$Mn
(R = La and Ce). The Mn ions present in `dilute' concentration of
just 3 molar percent form a sublattice with an unusually large
Mn-Mn near neighbor distance of $\sim$8.5 \AA. While the existence
of (RPd$_3$)$_8$M (where M is a p-block element) is already
documented in the literature, the present work reports for the
first time the formation of this phase with M being a 3d element.
In (LaPd$_3)_8$Mn, the Mn sub-lattice orders antiferromagnetically
as inferred from the peaks in low-field magnetization at 48 K and
23 K. The latter peak progressively shifts towards lower
temperatures in increasing magnetic field and disappears below 1.8
K in a field of $\sim$8 kOe. On the other hand in (CePd$_3)_8$Mn
the Mn sublattice undergoes a ferromagnetic transition around 35
K. The Ce ions form a dense Kondo-lattice and are in a
paramagnetic state at least down to 1.5 K. A strongly correlated
electronic ground state arising from Kondo effect is inferred from
the large extrapolated value of C/T = 275 mJ/Ce-mol K$^2$ at T = 0
K. In contrast, the interstitial alloys RPd$_3$Mn$_x$ ($x$ = 0.03
and 0.06), also synthesized for the first time, have a spin glass
ground state due to the random distribution of the Mn ions over
the available {\bf{1b}} sites in the parent RPd$_3$ crystal
lattice.
\end{abstract}

\maketitle

\section{\label{sec:level1}Introduction:\protect}% \\

The RPd$_3$ compounds have the AuCu$_3$-type crystal structure
with R and Pd atoms occupying the corners (\textbf{1a}) and face
centers (\textbf{3c}) of the cubic unit cell, respectively. A
unique way of alloying RPd$_3$ (R = La to Lu) compounds was
reported by Dhar et al., when they observed that smaller p-block
atoms like boron could be incorporated interstitially in the
RPd$_3$ crystal lattice\: \cite{dhar:mrb}.The smaller boron atoms
occupy the body-center (\textbf{1b}) position in the RPd$_3$ unit
cell forming a defect pervoskite-type structure with formula
RPd$_3$B$_x$ (0 $< x <$ 1). It was later found that silicon could
also be incorporated in the RPd$_3$ unit cell up to x $\sim$ 0.3\:
\cite{malik:ssc}. Alloying with boron and silicon resulted in the
expansion of the lattice and changed drastically the valence and
hence the magnetic properties of CePd$_3$ and EuPd$_3$\:
\cite{dhar:mrb,dhar:prb1,malik:ssc,dhar:jap,dhar:prb2}. Briefly,
cerium ions changed from valence fluctuating state in CePd$_3$ to
trivalent state in CePd$_3$B$_x$ for x $>$ 0.12, while the lattice
expansion forced the europium ions to change from the trivalent
state in EuPd$_3$ to the valence fluctuating state in
EuPd$_3$B$_x$ alloys, with possible charge ordering in EuPd$_3$B.\\

An interesting extension of the above-mentioned interstitial
alloying has recently been reported by Gordon and DiSalvo\:
\cite{gordon:naturf}. They found that alloying CePd$_3$ with an
appropriate amount of Sb leads to the formation of a cubic
superstructure, (CePd$_3$)$_8$Sb, closely related to the
AuCu$_3$-type structure of CePd$_3$. The large sized Sb atoms
deform the Pd octahedra, such that the available {\bf{1b}} sites
in the CePd$_3$ lattice are occupied by Sb atoms only in a regular
pattern forming a simple cubic sublattice with near neighbor
Sb$-$Sb separation slightly more than twice the lattice parameter
of the parent CePd$_3$ unit cell. In the superstructure the Ce
atom remains coordinated by 12 Pd atoms and the Ce$-$Ce separation
is increased by 2 to 3\% over its value in CePd$_3$. A rigorous
discussion of the superstructure can be found in\:
\cite{gordon:naturf,surjeet:jpcm}. Later, it was found that the
superstructure (CePd$_3$)$_8$M is also formed for other p-block
elements M = Ga, In, Sn, Pb and Bi, and for (LaPd$_3$)$_8$In\:
\cite{gordon:physb}. (CePd$_3$)$_8$M compounds order magnetically
below 10 K in contrast to the Pauli paramagnetic, valence
fluctuating ground state in CePd$_3$, and they exhibit
heavy-fermion like enhanced low temperature electronic specific
heat coefficient, $\gamma$\:
\cite{cho:prb,jones:physb,mitra:physb}. Further investigations
showed that (CePd$_3$)$_8$M (M = Al and Ge) also form, order
magnetically below 10 K and exhibit a dense Kondo lattice
behavior\: \cite{jones:physb,mitra:ssc,surjeet:jpcm}. Jones et al.
also tried to synthesise (CePd$_3$)$_8$M for M = Zn and Te
\cite{jones:physb}. Though their powder x-ray diffraction patterns
showed the signature of superstructure formation, there were some
doubts about the homogeneity and the exact stoichiometry of the
samples. The case for M = Zn is interesting in the sense that Zn
is not a p--block element and it motivated us to search for other
atoms in the 3d series that might lead to superstructure
formation. \\

In the present work, we report for the first time the formation of
`dilute-Mn' ternary compounds (RPd$_3$)$_8$Mn for R = La and Ce.
The word `dilute' emphasizes the fact that the Mn concentration in
these compounds is low (about few molar percent) and yet the Mn
ions form a periodic lattice with unusually large nearest neighbor
Mn$-$Mn separation allowing localization of the 3d moments.
Recently such `dilute-Mn' compounds have attracted considerable
attention. Negative colossal magnetoresistance has been reported
in the single crystals of Zintl compounds Eu$_{14}$MnSb$_{11}$ and
Eu$_{14}$MnBi$_{11}$, despite antiferromagnetic ordering in the
later compound \cite{zintl:shelton}. In Yb$_{14}$MnSb$_{11}$
(where Yb ions are divalent and hence non-magnetic) the localized
Mn$^{3+}$ moments are coupled via conduction electrons and order
ferromagnetically at 53 K \cite{zintl:chan,zintl:canfield}.
Yb$_{14}$MnBi$_{11}$ exhibits two magnetic transitions: a
ferromagnetic transition at 58 K followed by a second transition
at 28 K \cite{zintl:chan}. It was pointed out in ref.\:14 that
similar `dilute' intermetallic compounds may be a good starting
point to find 3d-Kondo lattice systems, analogous to the well
known and extensively studied 4f-Kondo lattice/heavy fermion
systems. We find that in (RPd$_3$)$_8$Mn (R = La, Ce) the nearest
neighbor Mn$-$Mn separation is unusually large $\sim8.5$ \AA. In
(LaPd$_3$)$_8$Mn the Mn ions are in a localized 2+ valence state
and order antiferromagnetically below $\sim$50 K. In
(CePd$_3$)$_8$Mn the Ce ions form a dense Kondo lattice and do not
order magnetically at least down to 1.5 K. However, the Mn
sublattice orders ferromagnetically near 35 K. In contrast the
alloys RPd$_3$Mn$_x$ (x = 0.03 and 0.06), also synthesized for the
first time, have a spin glass ground state due to the random
occupation of {\bf{1b}} sites by the Mn ions in the RPd$_3$ lattice.\\

\section{\label{sec:level2}Experiment\protect}% \\

Initially compounds of the composition (RPd$_3$)$_8$Mn for R = La
and Ce were made by the standard technique of arc melting in an
inert atmosphere of argon. First an ingot of RPd$_3$ was melted
and then Mn added to it by gently focusing the arc on the pellet
of RPd$_3$ so that Mn, which is volatile, is not directly heated.
The button of the ternary alloy was flipped over and re-melted
several times to ensure homogenization. The initial and final
weights of the pellets are practically the same and hence there is
no loss of Mn. We believe that the success of arc melting
technique for making these samples is partially due to the fact
that the molar percentage fraction of Mn is low ($\sim3$ \%).
Later, LaPd$_3$Mn$_x$ ($x$ = 0.03, 0.06) and CePd$_3$Mn$_x$ ($x$ =
0.02, 0.06) were also made by the same method. The samples were
checked for their phase purity by the standard powder x-ray
diffraction technique and optical micrography. The resistance
between 1.5$-$300\,K was measured by the four-probe DC method on a
rectangular parallelepiped piece of uniform cross-section.
Contacts with the sample were made using gold wire and silver
paste. The following method was used to remove the uncertainty in
the value of resistivity arising due to the error associated with
the measurement of distance between the voltage probes. The two
current and one voltage probes were fixed to the sample, the room
temperature resistance (R) of the sample was then measured as a
function of distance by moving the second voltage probe in steps
of ${\delta}L$ = 0.05mm. The slope of the R vs $L + {\delta}L$
(where $L$ is the initial distance between the moving and fixed
voltage probes; the slope is independent of the choice of L) is
used to normalize the value of resistivity obtained from silver
contact measurements. The heat capacity was measured by using the
semi-adiabatic, heat pulse method. Magnetization as a function of
temperature and field was recorded by using VSM (Oxford
Instruments) and SQUID (Quantum Design)
magnetometers.\\

\section{\label{sec:level3}Results and discussions\protect}% \\
\subsection{\label{sec:levelA}Powder X-ray diffraction patterns}
A representative powder x-ray diffraction pattern of
(LaPd$_3$)$_8$Mn obtained using Cu-K$\alpha$ radiations is shown
in Fig. 1. The crystal structure was refined using the Rietveld
refinement program FULLPROF\,\cite{fullprof} and an output of such
a fitting is also presented in Fig. 1. The Bragg peaks for the
powder pattern of (LaPd$_3$)$_8$Mn, generated by the FULLPROF, are
shown as two sets of vertical bars in Fig. 1. While the bigger and
thicker bars correspond to the reflections one would expect from a
face centered cubic LaPd$_3$ lattice, the smaller bars (labelled
as satellite peaks in the figure) correspond to additional low
intensity satellite peaks arising due to the periodic occupation
of {\bf{1b}} sites by the bigger Mn atoms resulting in the
formation of a cubic superstructure of size about twice the size
of the LaPd$_3$ unit cell. The powder diffraction pattern of
(CePd$_3$)$_8$Mn, similarly, showed the formation of cubic
superstructure. The lattice parameters \emph{a} of
(LaPd$_3$)$_8$Mn and (CePd$_3$)$_8$Mn are 8.466 and 8.415\:\AA\
respectively, which is also the nearest neighbor Mn$-$Mn
separation in the two compounds. The powder diffraction patterns
of RPd$_3$Mn$_x$ ($x=0.03$ and 0.06 for R = La and 0.02 and 0.06
for R = Ce) alloys, on the other hand, are similar to those of
RPd$_3$B$_x$ alloys: characterized by the absence of satellite
peaks and a slight lattice expansion over the parent RPd$_3$
lattice. We find that the superstructure in (RPd$_3$)$_8$Mn (R =
La and Ce) is sensitive to the heat treatment. For example,
annealing the (CePd$_3$)$_8$Mn samples at 900\:$^\circ$C
diminishes the low intensity satellite peaks and gives rise to new
peaks, indicating a degradation of the phase by the heat
treatment. All the data reported in this work
were taken on the as-cast samples. \\

\subsection{\label{sec:levelB}(LaPd$_3$)$_8$Mn}
LaPd$_3$ is diamagnetic at room temperature and shows a nearly
temperature independent magnetic susceptibility\:
\cite{scoboria:jap}. The inverse molar susceptibility,
$\chi^{-1}$, of (LaPd$_3$)$_8$Mn measured in an applied field of
3\,kOe is shown in Fig. 2.  Between 100 and 300 K the data can be
fitted to a modified Curie-Weiss expression: $\chi = \chi{_0}$ +
C/(T - $\Theta_p)$ where C is the Curie constant and $\chi_0$ is
the temperature-independent contribution to the total magnetic
susceptibility and $\Theta_p$ is the paramagnetic Curie
temperature. The values of the best-fit parameters are:
$\mu_{eff}$ = 5.9 $\mu_B$/f.u. (derived from C), $\Theta_p$ = 52.1
\!K and $\chi_0$ = $-$0.003 emu/mol. The $\mu_{eff}$ is comparable
to the theoretical value of 5.92 \!$\mu_B$ for Mn$^{2+}$ (S = 5/2)
ion and suggest that the Mn ions in this compound are in a 3d$^5$
electronic configuration. The negative $\chi_0$ (about 20\% of the
total susceptibility at 300 K) arises from the the diamagnetic
contribution of the LaPd$_3$ network and is comparable to the
susceptibility of (LaPd$_3$)$_8$In which is reported to be
$-1.06\times10^{-3}$ emu/mol at 293 K\: \cite{gordon:physb}. The
susceptibility of (LaPd$_3$)$_8$Mn measured in a field of 1 kOe is
shown in the inset of Fig. 2. At low temperatures the
susceptibility shows a peak at 48 \!K (T1) indicating an
antiferromagnetic transition. In addition there is a smaller peak
at 23 \!K (T2) indicative of a second phase transition.\\

Fig. 3a shows that the magnetization of (LaPd$_3$)$_8$Mn measured
in zero-field-cooled (ZFC) and field-cooled (FC) mode in an
applied field of 100 Oe is practically the same. Further, we do
not observe any frequency dependence of the AC susceptibility
measured at 1, 9 and 99 Hz (Fig. 3b). These data show conclusively
that the peaks in the magnetization arise due to the long range
magnetic order and not due to any spin-glass-type freezing of the
Mn moments. The magnetization, [M(T)/H]$_H$, of (LaPd$_3$)$_8$Mn,
measured in several applied fields below 60 K is depicted in Fig.
4. [M(T)/H]$_H$ in the ordered state exhibits very unusual
behavior in relatively low applied fields. Noteworthy is the shift
of T2-peak to lower temperatures: above $\sim3$ kOe the T2-peak
shifts rapidly to lower temperatures and disappears below 1.8 K in
an applied field of $\sim8.25$ kOe. The magnitude of M/H, on the
other hand, increases with the increase in the magnetic field.\\

The isothermal magnetization [M(H)]$_T$ measured at several
temperatures between 2K and 50K is shown in Fig. 5. For each T,
[M(H)]$_T$ is measured by cooling the sample from above 60 K to
the desired temperature in a nominal zero field. [M(H)]$_{1.8 K}$
undergoes a sharp metamagnetic transition in a `critical field' of
$\sim$\,8.25 \!kOe, at which the T2-peak of magnetization
disappears as seen in Fig. 4. At higher temperatures the
metamagnetic transition becomes less sharper and the corresponding
field at which the matamagnetic transition occurs also decreases.
The linearity of the magnetization at 50 K characterizes the
response in the paramagnetic state. [M(H)]$_{2K}$ (shown as inset
in Fig. 5 up to 120 kOe) has a tendency towards saturation
attaining a value of $\mu_s$ = 4.6 $\mu_B$/Mn in 120 kOe (close to
the theoretical value of 5$\mu_B$ for Mn$^{2+}$ ion). From the
values obtained for $\mu_{eff}$ and $\mu_s$ we conclude that the
3d state of the Mn ions in (LaPd$_3$)$_8$Mn has a localized
character in contrast to the usual band like character of Mn 3d
state in the
intermetallics.\\

It may be noticed that though $\Theta_p$ obtained from the high
temperature Curie-Weiss behavior of the susceptibility data is
positive, indicative of ferromagnetic correlations, the peak in
$\chi$ at T1 and the nature of isothermal magnetization suggest
antiferromagnetic ordering. The data presented in Fig. 4 and 5
shows that the antiferromagnetic state is not robust. We speculate
that the ferromagnetic and antiferromagnetic interactions are
competing in this compound. It is worthy of mention here that the
Mn sublattice in (CePd$_3$)$_8$Mn orders ferromagnetically (see
below) and our preliminary investigations reveal ferromagnetic
ground state in (LaPd$_{2.75}$Rh$_{0.25}$)$_8$Mn and
(LaPd$_{2.75}$Ag$_{0.25}$)$_8$Mn. Thus it appears that the nature
of magnetic ordering is rather sensitively dependent on slight
variations in lattice parameter, conduction electron
concentration, etc.\\

It is interesting to note here that the  metamagnetic transitions
observed in (LaPd$_3$)$_8$Mn is qualitatively similar to that
observed in bilayer itinerant metamagnet Sr$_3$Ru$_2$O$_7$ where
the quantum critical fluctuations associated with the
metamagnetism at very low temperatures has been shown to give rise
to a non-Fermi liquid behavior in
magnetotransport\,\cite{perry:prl,Grigera:science}. Therefore, it
will be very interesting and pertinent to investigate transport
and magnetic properties of (LaPd$_3$)$_8$Mn at very low
temperatures and in fields close to the `critical field'
$\sim8.25$ kOe: the field at which the T2-peak in magnetization
disappears below 1.8 K or the isothermal magnetization exhibits
metamagnetic transition at 1.8 K. We present here our preliminary
results on the magnetotransport measurements of (LaPd$_3$)$_8$Mn
down to 1.5 K.\\

The zero-field resistivity, $\rho$(H\,=\,0, T), data of
(LaPd$_3$)$_8$Mn between 1.5$-$300 K are shown in the inset of
Fig. 6. A rather low residual resistivity of $\sim7\mu\Omega$\,cm
and a moderately good residual resistivity ratio of about 4.5 is
observed. At high temperatures $\rho$(0, T) decreases almost
linearly with temperature down to 100 K. The magnetic transition
at T1 is clearly reflected by a distinct change in the slope
around 48 K below which $\rho$(0, T) drops precipitously due to
the loss of spin-disorder scattering. Around T2 = 23 K the
zero-field resistivity is smooth with no discernible anomaly. The
magnetoresistivty (MR) measured at few selected fields (Fig. 6)
shows interesting features, which correlate nicely with the
changes in magnetization in applied fields. The MR, measured in
applied fields of 7 and 10 \!kOe, is positive between the
temperatures corresponding to T1 and T2-peaks in the low field
magnetization, in accordance with the positive MR typically
observed in the antiferromagnetic state. However, at lower
temperatures the MR changes sign and exhibits a field dependent
peak, the peak position T2 coinciding with that in the DC
magnetization at the same value of the applied field. The low
temperature peak in MR vanishes for fields above $\sim$\,8.25
\!kOe within the temperature range of our measurements, thus
mimicking the corresponding behavior seen in the DC magnetization.
Thus, while the zero-field resistivity is apparently featureless
at T2 = 23 K, the in-field resistivity changes distinctly at
(field dependent) T2 pointing at the bulk nature of changes taking
place at T2.

The in-field resistivity measured at selected fields are plotted
in Fig. 7a as a function of T$^2$. The zero-field resistivity and
the resistivity in applied fields greater than the `critical
field' (namely 8.5, 10 and 15 \!kOe) follows a T$^2$ variation
between 1.5 and 10 K (T$^2$ behavior in zero-field persists up to
nearly 15 K). The data in 10 and 15 kOe are nearly coincident with
the 8.5 kOe data and are, therefore, not shown in the figure.
Evidently, for applied fields less than the `critical field'
(namely 5, 7.8 and 8 kOe) a similar quadratic dependence of
resistivity is not observed at low temperatures. However, at
temperatures above the field dependent T2-peak the data nearly
coincides with the data taken in 8.5 \!kOe and thus follow a T$^2$
behavior. MR at a fixed temperature of 1.6 \!K was measured up to
60 \!kOe and is shown in Fig. 7b. MR drops precipitously to
$-10\%$ at $\sim$\,8.25 \!kOe , correlating with the metamagnetic
transition (Fig. 5). In Sr$_2$Ru$_3$O$_7$ also the metamagnetism
is similarly observed in the MR\,\cite{perry:prl}. In higher
fields the MR becomes less negative attaining a value of $-3\%$ at
60 \!kOe. It would be interesting to study the thermal variation
of the magnetoresistivity in magnetic field near the
`critical field' at very low temperatures for possible non-Fermi
liquid behavior.\\

The heat capacity, C, of (LaPd$_3$)$_8$Mn, shown in Fig. 8
increases monotonically with temperature up to 60 K. No
discernible anomaly is seen at T1 and T2 where the Mn sub-lattice
undergoes transitions. Recalling that the heat capacity at a bulk
magnetic transition for spin S = 1/2 shows a peak of 25 J/mol K in
the mean-field approximation, the lack of anomaly at first appears
puzzling. However, it should be noticed that in (LaPd$_3$)$_8$Mn
the Mn ions are just about 3 molar percent. Therefore, the total
heat capacity at high temperatures is predominantly due to the
phonons. The background lattice heat capacity masks the anomaly
expected at the magnetic transitions near T1 and T2. The heat
capacity of iso-structural non-magnetic analogue (LaPd$_3$)$_8$Ga
is also shown in Fig. 8. Assuming that the conduction electron
density of states and the phonon contributions in (LaPd$_3$)$_8$Mn
are exactly the same as in iso-structural (LaPd$_3$)$_8$Ga, we
have extracted the excess magnetic contribution, C$_{mag}$, by
subtracting the two heat capacities, which is shown in the upper
inset of Fig. 8. The relatively poor quality of C$_{mag}$ is easy
to understand. Typically, our heat capacity set up has accuracy to
within 3$-$4\%. Subtracting the heat capacities of the two
compounds will give large statistical errors ($\sim$ 10$-$20 J/mol
K) at high temperatures (C = 250 J/mol K at $\sim$ 50 K). A peak
in C$_{mag}$ centered at 48 K ( = T1) is indeed seen though no
apparent anomaly is observed at T2 as in the zero-field
resistivity. In the low temperature region (3 $<$ T $<$6 K), C/T
vs. T$^2$ of (LaPd$_3$)$_8$Mn shows linear behavior (see lower
inset of Fig. 8) with least square fitted values of $\gamma$ =
57.8$\:\pm\:$1.6 mJ/mol K$^2$ and $\beta$ = 6.75$\:\pm\:$0.1
mJ/mol K$^4$, where $\gamma$ and $\beta$ have their usual meaning.
For antiferromagnets the spin wave contribution varies as T$^3$
and should not therefore affect the value of $\gamma$. For the
non-magnetic analogue (LaPd$_3$)$_8$Ga $\gamma$ = 2.3$\:\pm\:$2.1
mJ/ mol K$^2$ which is comparable to 2.21 mJ/mol K$^2$ in
(LaPd$_3$)$_8$In reported by Cho et al.\: \cite{cho:prb}. The data
between 3 and 10 K were least square fitted by including
additional T$^4$ and T$^6$ terms in the expression for C/T. The
values of $\gamma$ thus obtained are 61.2$\:\pm\:$5.3 and
5.4$\:\pm\:$3.3 mJ/mol K$^2$ for Mn and Ga compounds respectively.
Within the limits of error the two sets of values are comparable
for each compound. The value of $\gamma$ for (LaPd$_3$)$_8$Mn is
an order of magnitude larger than the corresponding value for
(LaPd$_3$)$_8$Ga and shows that there is an enhancement in the
electronic density of states at the Fermi
level in the Mn compound.\\

\subsection{\label{sec:levelC}LaPd$_3$Mn$_x$ ($x = 0.03, 0.06$)}
It was inferred above from the powder x-ray diffraction pattern of
that the Mn ions in LaPd$_3$Mn$_x$ ($x$ = 0.03 and 0.06) are
randomly distributed over the {\bf{1b}} sites in LaPd$_3$ lattice.
A random distribution of Mn-ions coupled via RKKY exchange will
give rise to a spin-glass type of freezing of the Mn moments as
observed, for example, in the canonical spin-glasses like
{\bf{Cu}}$-$Mn, {\bf{Au}}$-$Fe, etc\: \cite{mydosh:sg}. Indeed,
our data show convincingly the spin glass state for $x$ = 0.03 and
0.06 in LaPd$_3$Mn$_x$. Fig. 9a shows the susceptibility of
LaPd$_3$Mn$_{0.03}$ measured in an applied field of 50 Oe both in
ZFC and FC mode. In the ZFC mode the susceptibility exhibits a
peak at 2.8 K below which large thermomagnetic irreversibility is
observed between ZFC and FC data. These features are typically
associated with a spin-glass freezing of the magnetic moments.
This is further corroborated by the frequency dependent peak
position of the ac-susceptibility measured at 1, 9 and 999 Hz as
shown in Fig. 9b. The peak in the AC-$\chi$ shifts to higher
temperatures with the increase in the frequency of the AC
excitation. The corresponding data for LaPd$_3$Mn$_{0.06}$ are
shown in Fig. 10a and 10b. A similar behavior is seen but with a
freezing temperature (T$_f$) of 5.5 K. The ratio
$\Delta$T$_f$/$\:$T$_f$[$\Delta$($\log \omega$)], a quantitative
measure of the frequency-dependent peak shift, for $x$ = 0.03 and
0.06 is 0.010 and 0.0098 respectively, which compares well with
the values for the canonical spin glasses (0.005 and 0.01 for
{\bf{Cu}}$-$Mn and {\bf{Au}}$-$Fe, respectively). This ratio
offers a good criterion for distinguishing a canonical spin glass
from a spin-glass-like material, superparamagnets for example,
where it is often larger by about an order of magnitude\:
\cite{mydosh:sg}. The inverse molar susceptibility of
LaPd$_3$Mn$_{0.03}$ and LaPd$_3$Mn$_{0.06}$ measured up to 300 K
is shown in Fig. 11. As above, the data are least squares fitted
to the modified Curie-Weiss expression in the range 100 to 300 K.
The values of the best-fit parameters are: $\mu_{eff}$ = 5.85
$\mu_B$/Mn and 5.70 $\mu_B$/Mn, $\Theta_p$ = 15.8 K and 32.3 K and
$\chi_0$ = $-0.0002$ emu/mol and $-0.0002$ emu/mol respectively.
$\mu_{eff}$ in both the alloys is comparable to the theoretical
value of 5.92 $\mu_B$ for Mn$^{2+}$ (S = 5/2) ion. Thus the
magnetic moments of the Mn-ions in these two alloys
freeze randomly at T$_f$ due to the random RKKY interaction.\\

In canoniocal spin-glasses like {$\bf{Cu}$}$-$Mn the magnetic
contribution to the heat capacity, C$_{mag}$, exhibits a broad
maximum above T$_f$ and gradually falls off at increasing
temperatures\,\cite{mydosh:sg}. Fig. 12a and 12b shows the heat
capacity of LaPd$_3$Mn$_x$ ($x$ = 0.03, 0.06). The heat capacity
of non-magnetic reference LaPd$_3$ is also displayed in the
figure. The insets show C$_{mag}$ obtained by subtracting the heat
capacity of LaPd$_3$ from LaPd$_3$Mn$_x$ alloys. Extraction of
C$_{mag}$ form the total heat capacity is an arduous task
particularly at high temperatures where the errors associated with
the absolute values of the total heat capacity may exceed the
magnetic contribution from the Mn-moments. For $x$ = 0.03 a broad
maximum of height 120 mJ/mol K in C$_{mag}$ is observed at 4.5 K
giving additional evidence of a spin glass freezing of Mn moments
in LaPd$_3$Mn$_{0.03}$. For $x$ = 0.06, C$_{mag}$ has large
scatter above 15 K nonetheless indicates a  broad peak
centered at 16 K.\\

\subsection{\label{sec:levelC}(CePd$_3$)$_8$Mn}
CePd$_3$ is an archetypal valence fluctuating compound with
enhanced Pauli-paramagnetic ground state ($\chi\sim$\:10$^{-3}$
emu/mol and $\gamma$ = 50 mJ/mol K$^2$ as T $\rightarrow$ 0 K)\,
\cite{gardner:jphys,besnus:jphys,wohllenben:prl,wohllenben:prb}.
Fig. 13a shows the ZFC and the FC magnetization of
(CePd$_3$)$_8$Mn in an applied field of 1 kOe. The magnetization
increases sharply below 50 K like in a ferromagnetic transition
and exhibits a broad peak at 20 K in the ZFC mode. There is a
pronounced difference between the ZFC and FC magnetization below
the peak temperature. The field dependence of magnetization at 2.8
K, shown in the inset of Fig. 13c, is characterized by both
hysteresis and a high coercive field of 6 kOe. N\'{e}el has shown
that the magnetization M(T) of a ferromagnet below its Curie
temperature, T$_c$, in applied field h much smaller than the
coercive field h$_c$(T) is approximately given by the relation
M(T) $\approx$ m$_s$(T)\,h$^2$/h$_{c}^2$(T), where m$_s$(T) is the
saturation magnetization\: \cite{neel:annphys}. Therefore, if
below T$_c$ the anisotropy or the coercive field builds up at a
rate much faster than m$_s$(T) then M(T) should show a peak at
some temperature. We believe the broad peak at 20 K in ZFC data
arises due to the appreciable coercivity/anisotropy in this
ferromagnetic material. The FC data in such cases typically show a
ferromagnetic saturation below the peak but we find that in
(CePd$_3$)$_8$Mn the FC magnetization decreases by about 5 \%
below 12 K; the reasons for this are not clear to us. The
susceptibility of (CePd$_3$)$_8$Mn measured in an applied field of
5 kOe is shown in Fig. 13b. The data above 75 K can be
satisfactorily fitted to the modified Curie-Weiss law. The
best-fit parameters obtained from such a fitting are: $\chi_0$ =
0.002 emu/f.u., $\mu_{eff}^{observed}$ = 7.15 $\mu_B$/f.u. and
$\theta_p = +38.7$ K. The positive value of $\theta_p$ indicates
ferromagnetic interactions. Indeed the Mn sublattice in
(CePd$_3$)$_8$Mn orders ferromagnetically around 32 K (refer to
Arott plot analysis and heat capacity section below). To deduce
the effective magnetic moment per Ce ion it is reasonable to
assume that the valence of Mn in (CePd$_3$)$_8$Mn is same as
observed in the isomorphous (LaPd$_3$)$_8$Mn. We further assume
that in the paramagnetic regime the contribution of Ce and Mn to
the observed effective magnetic moment is simply additive and,
therefore, it can be expressed as: $\mu_{eff}^{observed}$ =
$\sqrt{8(\mu_{eff}^{Ce})^2 + (\mu_{eff}^{Mn})^2}$, where
$\mu_{eff}^{Ce}$ and $\mu_{eff}^{Mn}$ are effective magnetic
moments per Ce and Mn ions respectively in (CePd$_3$)$_8$Mn. Since
in (LaPd$_3$)$_8$Mn the effective magnetic moment per Mn ion is
$5.9 \mu_B$, using this relation we get $\mu_{eff}^{Ce} = 2.42
\mu_B$ per Ce ion, which is close to the theoretical value of 2.54
$\mu_B$ for a free Ce$^{3+}$ ion. Indicating that the valence of
Ce in (CePd$_3$)$_8$Mn is close to +3.\\

Fig. 14a shows [M(H)]$_T$ at various temperatures. In Fig.14b we
have replotted the data as H/M vs.\,M$^2$ at T = 20, 30, 40 and 45
K. Such plots, known as Arrott plots, are useful in determining
the Curie temperature of a ferromagnet. For a ferromagnet the
Arrott plot at T$_c$ should pass through the origin while the
plots for T $<$ T$_C$ will cut the horizontal axis at a non-zero
value, which is the square of the spontaneous magnetization of the
ferromagnet (M$_s$). The plots for T $>$ T$_C$ do not intercept
the horizontal axis, since the spontaneous magnetization is zero.
Following the standard procedure (see for example Ref. 26)
\cite{sullow:prb}, the high-field (higher than 5 kOe) data for
each temperature is fitted to a fourth order polynomial in M$^2$
which is extrapolated to determine M$^{2}_{s}$. We find M$_s$ at
40 K is zero while it has a non-zero value at 30 K, indicating
that the ferromagnetic ordering temperature of (CePd$_3$)$_8$Mn
lies between these two temperatures. Isothermal magnetization data
at few more temperatures between 30 and 40 K are required to
locate T$_C$ exactly. The magnetization at 2.8 K does not saturate
even in 120 kOe and its field variation suggests a component
varying almost linearly with the field. We believe that the linear
component arises from the contribution due to the Ce ions which
are in the paramagnetic state down to, at least, 1.5 K (see below,
heat capacity and resistivity data). We thus attribute the
ferromagnetic transition in (CePd$_3$)$_8$Mn to the Mn sublattice.
The magnetization at 2.8 K in 120 kOe is 6.4 $\mu_B$/f.u. and
exceeds the corresponding value in (LaPd$_3$)$_8$Mn substantially
due to the (paramagnetic)
contribution from the Ce sub-lattice.\\

The resistivity ($\rho$) of (CePd$_3$)$_8$Mn (Fig. 15) is about
four times larger than that of (LaPd$_3$)$_8$Mn at 300 K and its
thermal variation is qualitatively similar to that seen in the
dense Kondo lattice compounds. The relatively higher $\rho$ in
(CePd$_3$)$_8$Mn arises from the incoherent Kondo spin disorder
scattering. The peak at 16 K is not due to any phase transition as
both the heat capacity (see below) and magnetization data do not
show any anomaly at that temperature, but it is due to the
transition from the incoherent to coherent Kondo scattering regime
as seen, for example, in archetypal Kondo lattice compounds
CeAl$_3$, CeCu$_6$\: \cite{stewart:rmp} and also in Kondo lattice
antiferromagnet (CePd$_3$)$_8$M (M = Ga, Ge and Al)\:
\cite{mitra:ssc,gordon:physb,surjeet:jpcm}. The resistivity does
not apparently exhibit any anomaly at T$_C$ $\sim$ 35\,K because
its large negative temperature coefficient masks the decrease in
resistivity occurring due to the ordering of Mn ions (3
mol-percent). It is interesting to recall that the resistivity of
CePd$_3$ also shows a negative temperature coefficient and peaks
around 130\,K\: \cite{besnus:jphys}. Qualitatively the peak
temperature varies inversely with the characteristic Kondo/spin
fluctuation temperature. Therefore, there is a significant
reduction of the Kondo temperature in (CePd$_3$)$_8$Mn. However,
compared to Kondo-lattices (CePd$_3$)$_8$M (M = Ga, Ge, In, Sn, Sb
Bi and Al)\: \cite{cho:prb,mitra:physb,jones:physb, surjeet:jpcm}
which all order antiferromagnetically between 1.5 and 8 K, the
RKKY interaction is not strong enough to induce magnetic ordering
in (CePd$_3$)$_8$Mn.\\

The heat capacity of (CePd$_3$)$_8$Mn though comparable to that of
the La analogue is however larger in the entire temperature range
of 1.5 K to 60 K. This is understandable as the predominant (at
higher temperatures) lattice contribution will be nearly similar
in the two compounds (the atomic masses of La and Ce differ by
less than 1 \%) but there will be an additional 4f-derived
contribution, C$_{4f}$ due to the Ce-ions in (CePd$_3$)$_8$Mn. The
magnetic contribution to the heat capacity, C$_{mag}$ (which is
predominantly C$_{4f}$ considering that the Mn concentration is
low), was derived by subtracting the heat capacity of
(LaPd$_3$)$_8$Ga from (CePd$_3$)$_8$Mn. C$_{mag}$ displayed in
Fig. 16a exhibit a peak at $\sim$32 K which we attribute to the
ferromagnetic ordering of the Mn sub-lattice. The peak position of
C$_{mag}$ is in agreement with the estimate of the
Curie-temperature derived from Arrott plots. Inset of Fig. 16
shows C$_{mag}$/T vs. T$^2$ below 5 K. A linear extrapolation to T
= 0 K gives an intercept of 2.2 J/mol K$^2$ (275 mJ/Ce-mol K$^2$).
In the case of Ce compounds the screening of Ce magnetic moments
by the Kondo interaction results in an enhanced coefficient of the
electronic specific heat $\gamma$. The large value of C$_{mag}$/T
at T = 0 K shows that (CePd$_3$)$_8$Mn is a dense Kondo lattice.
We may recall that $\gamma$ of CePd$_3$ is $\sim$\,50mJ/mol K2.
The hybridization of Ce 4f-states with the conduction electrons is
reduced in the ternary compound thereby leading to a narrow band
density of states and an enhanced effective mass. There is no
signature of the magnetic ordering of Ce ions in (CePd$_3$)$_8$Mn
unlike (CePd$_3$)$_8$M (M = Ga, Ge, In, Sn, Sb Bi and Al)\:
\cite{mitra:ssc,gordon:physb,surjeet:jpcm} where the heat capacity
exhibits huge peaks between 1.5$-$10 K.\\

\subsection{\label{sec:levelD}CePd$_3$Mn$_x$ (x = 0.02, 0.06)}

The ZFC and FC magnetization data of CePd$_3$Mn$_{0.06}$ in 100 Oe
and the AC susceptibility measured at 9, 99 and 999 Hz are
displayed in Fig. 17a and 17b respectively. A peak in the DC
magnetization at 3.9 K with a pronounced thermomagnetic
irreversibility below the peak temperature and a frequency
dependent peak shift in  the AC susceptibility confirms the
occurrence of a spin-glass freezing in CePd$_3$Mn$_{0.06}$ at 3.9
K. Since there are two magnetic entities (Ce and Mn) the
magnetization data do not distinguish which one of these is
getting frozen at 3.9 K. To clarify the situation we measured the
heat capacity of CePd$_3$Mn$_{0.06}$ down to 1.5 K as shown in the
inset of Fig. 18. The heat capacity data are featureless (with the
exception of a minor anomaly at 6.2 K due to the presence of
cerium oxide in the sample). The magnetic contribution, C$_{mag}$,
is obtained as C$_{mag}$(T) = [C$_{(CePd_3Mn_{0.06})}$(T) -
C$_{(LaPd_3)}$(T)] and is shown in the main figure (Fig. 18). A
broad peak in C$_{mag}$ centered at T $\sim 2\,$T$_{f}$, with a
peak height of $\sim 1$ J/mol K corroborates the spin glass
freezing. A rough estimate of the magnetic entropy released up to
20 K yields S$_{mag}$ $\sim 1.9$ J/mol K. This value is nearly
$1/3$ of what one would expect from the spin-glass freezing of one
mol of Ce ions. These two features: absence of a broad peak in the
C and S$_{mag}$ $\sim 1.9$ J/mol K rule out the spin-glass
freezing of the Ce ions at 3.9 K. To conclude, in
CePd$_3$Mn$_{0.06}$ the Mn ions freeze in a spin-glass state below
3.9 K, while the Ce ions remain paramagnetic down to the lowest
temperature measurements.\\

The resistivity of CePd$_3$Mn$_{0.02}$ and CePd$_3$Mn$_{0.06}$ is
shown in Fig. 15. For x = 0.02 the resistivity is qualitatively
similar to that of CePd$_3$, the peak having shifted from 130 K in
CePd$_3$ to $\sim$\,100 K indicating that the addition of Mn at
this level does not influence the electronic state of Ce-ions to
any considerable extent. But for x = 0.06 which further expands
the lattice, the resistivity shows a negative d$\rho$/dT down to
the lowest temperature of 1.5 K. For single ion Kondo effect the
magnetic part of the resistivity saturates as T$^2$ at low
temperatures for T $<<$ T$_K$, where T$_K$ is the Kondo
temperature. In CePd$_3$Ga$_{0.05}$, for example, the T$^2$
dependence extends up to T = 20 K\: \cite{cepd3gax}. However our
low temperature data do not follow a similar behavior. For example
between 1.7 - 4 K a least square fitting of the data to the
expression $\rho = \rho_{0}$ - AT$^n$ gives $\rho_0$= 344.9
$\mu\Omega$ cm, A =0.065$\mu\Omega$ cm K$^{-2}$ and n =1.7.
Absence of T$^2$ behavior in the low temperature resistivity of
anomalous f-electron systems has been interpreted to arise from a
non-Fermi liquid ground state. Since the low temperature
resistivity in CePd$_3$Mn$_{0.06}$ includes the contribution
arising from the spin glass freezing of the Mn moments at 3.9 K we
cannot infer conclusively a non-Fermi liquid ground state in this
alloy.

\section{\label{sec:level3}Summary\protect}% \\
We have synthesized and studied the magnetic, thermodynamic and
magnetotransport properties of two new interstitial `dilute-Mn'
compounds (LaPd$_3$)$_8$Mn and (CePd$_3$)$_8$Mn. Interstitial
alloys CePd$_3$Mn$_{x}$ ($x = 0.02, 0.06$) and LaPd$_3$Mn$_x$ ($x
= 0.03, 0.06$) were also studied to bring out the differences in
the properties of the compounds and the alloys. The Mn ions form a
regular sub-lattice in the compounds and undergo long range
magnetic ordering, while they are randomly distributed in the
alloys and show spin glass behavior. The peculiarity of the
antiferromagnetic state in (LaPd$_3$)$_8$Mn is reflected in the
occurrence of two peaks (at 48 and 23 K) in its low field
magnetization and an unusually strong field-induced shift of the
lower peak (T2) towards T = 0 K at relatively low fields.
Concomitantly, a metamagnetic transition is seen in the
magnetization in the ordered state, which is also clearly
reflected in the magnetotransport data. The magnetic ordering due
to Mn-sub-lattice is ferromagnetic in (CePd$_3$)$_8$Mn, though the
Ce sub-lattice shows dense Kondo lattice behavior and remains
paramagnetic at least down to 1.5 K. It would be interesting to
probe the magnetic behavior of (LaPd$_3$)$_8$Mn in the vicinity of
the metamagnetic field at very low temperatures.

\begin{acknowledgments}
We thank Dr. Pratap Raychaudhuri for useful discussions.
\end{acknowledgments}
\newpage

\bibliography{R8P24Mnx}% Produces the bibliography via BibTeX.

\newpage

\begin{description}
\item [Figure Captions:]

 \item [Fig. 1.] The observed (cross) and fitted (solid
line) powder x-ray diffraction pattern of (LaPd$_3$)$_8$Mn. The
vertical bars indicate the position of the Bragg peaks.

\item [Fig. 2.] The inverse molar susceptibilities $\chi^{-1}$ and
$(\chi -\chi_0)^{-1}$ of (LaPd$_3$)$_{8}$Mn as a function of
temperature. $\chi_0$ is estimated by fitting the susceptibility
data to the modified Curie-Weiss expression (refer text for
details). The line is the best-fit of the data to the modified
Curie-Weiss expression.

\item [Fig. 3] The magnetic susceptibility of (LaPd$_3$)$_8$Mn as
a function of temperature: {\bf{(a)}} DC susceptibility ($\chi$)
in zero-field-cooled (ZFC) and field-cooled (FC) modes in an
applied field of 100 Oe. {\bf{(b)}} AC susceptibility ($\chi'$) at
1, 9 and 99 Hz.

\item [Fig. 4.] The magnetization (M/H) of (LaPd$_3$)$_8$Mn as a
function of temperature for a number of fixed applied fields. The
dotted line (guide to the eyes) trace the locus of the T2-peak
which shift towards lower temperatures with increase in the
applied magnetic field.

\item [Fig. 5.] The magnetization isotherms of (LaPd$_3$)$_8$Mn as
a function of applied magnetic field at several temperatures.
Inset: data plotted up to 120 kOe at 2K.

\item [Fig. 6.] The magnetoresistance (MR) of (LaPd$_3$)$_8$Mn is
shown as a function of temperature for several fixed applied
fields. The arrows (pointing upward) mark the position of T2-peak
which shift towards lower temperatures with increasing applied
fields. Inset: zero-field resistivity as a function of temperature
up to 300 K.

\item [Fig. 7.] \textbf{(a)} The in-field resistivity ($\rho$) of
(LaPd$_3$)$_8$Mn as a function of T$^2$ for several fixed applied
fields. Lines indicate a quadratic behavior of $\rho$ in T.
\textbf{(b)} The magnetoresistance (MR) of (LaPd$_3$)$_8$Mn as a
function of applied magnetic field at 1.6 K.

\item [Fig. 8.] The heat capacity (C) of (LaPd$_3$)$_8$Mn and its
non-magnetic analogue (LaPd$_3$)$_8$Ga shown as a function of
temperature. Upper inset: magnetic part (C$_{mag}$) of the heat
capacity in (LaPd$_3$)$_8$Mn. Lower inset: C/T plotted against
T$^2$ for the two compounds. The lines are linear extrapolation of
the data to T = 0 K.

\item [Fig. 9.] The magnetic susceptibility of LaPd$_3$Mn$_{0.03}$
as a function of temperature: \textbf{(a)} DC susceptibility
($\chi$) in zero-field-cooled (ZFC) and field-cooled (FC) modes in
an applied field of 100 Oe. \textbf{(b)} AC susceptibility
($\chi'$) at 1, 9 and 999 Hz. The continuous line is guide to the
eyes.

\item [Fig. 10.] The magnetic susceptibility of
LaPd$_3$Mn$_{0.06}$ plotted as a function of temperature:
{\bf{(a)}} DC susceptibility ($\chi$) in zero-field-cooled (ZFC)
and field-cooled (FC) modes in an applied field of 50 Oe.
{\bf{(b)}} AC susceptibility ($\chi'$) at 1, 9, 99 and 999 Hz. The
solid lines are guide to the eyes.

\item [Fig. 11.] The inverse molar susceptibilities: $\chi^{-1}$
and $(\chi -\chi_0)^{-1}$ of LaPd$_3$Mn$_{0.03}$ and
LaPd$_3$Mn$_{0.06}$ as a function of temperature. $\chi_0$ is
estimated by fitting the susceptibility data to the modified
Curie-Weiss expression (refer text for details). The solid lines
are the best-fit of the data to the modified Curie-Weiss
expression.

\item [Fig. 12.] \textbf{(a)} The heat capacity (C) of
LaPd$_3$Mn$_{0.03}$ plotted as a function of temperature. Inset:
The magnetic part in the heat capacity of LaPd$_3$Mn$_{0.03}$
(C$_{mag}$). \textbf{(b)} C and C$_{mag}$ (inset) in
LaPd$_3$Mn$_{0.06}$. C of the non-magnetic analogue LaPd$_3$ is
also shown both \textbf{a} and \textbf{b}.

\item [Fig. 13.] \textbf{(a)} The magnetization (M/H) of
(CePd$_3$)$_8$Mn in an applied field of 1 kOe as a function of
temperature in zero-field-cooled (ZFC) and field-cooled (FC) mode.
Inset: an exaggerated view of the low temperature maximum in M/H.
\textbf{(b)} The inverse molar susceptibility in an applied field
of 5 kOe as a function of temperature. The solid line is best-fit
of the data to the modified Curie-Weiss expression \textbf{(c)}
The hysteresis loop in the isothermal magnetization (M) of
(CePd$_3$)$_8$Mn measured at 2.8 K as a function of applied
magnetic field.

\item [Fig. 14.] The magnetization isotherms of (CePd$_3$)$_8$Mn
at several temperatures between 20 and 75 K are shown as a
function of temperature. \textbf{(b)} Arrot plots at 20, 30, 40
and 45 K in (CePd$_3$)$_8$Mn. The continuous lines are polynomial
(in M$^2$) fits to the high field data (please see text for
details).

\item [Fig. 15.] The resistivity of (CePd$_3$)$_8$Mn and
CePd$_3$Mn$_x$ ($x = 0.02, 0.06$) as a function of temperature.
Inset: an enlarged view of the same in low temperatures.

\item [Fig. 16.] \textbf{(a)} The magnetic part (C$_{mag}$) in the
heat capacity of (CePd$_3$)$_8$Mn is shown as a function of
temperature between 15 and 45 K. Inset: heat capacity (C) of
(CePd$_3$)$_8$Mn is shown between 1.5K and 50 K. \textbf{(b)} C/T
against T$^2$ at low temperatures. The line is linear
extrapolation of the data to T = 0 K.

\item [Fig. 17.] The magnetic susceptibility of
CePd$_3$Mn$_{0.06}$ as a function of temperature: {\bf{(a)}} DC
susceptibility ($\chi$) in zero-field-cooled (ZFC) and
field-cooled (FC) modes in an applied field of 100 Oe. {\bf{(b)}}
AC susceptibility ($\chi'$) at 9, 99 and 999 Hz. The solid line is
guide to the eyes.

\item [Fig. 18.] The magnetic part, Cmag, (left axis) and the
total heat capacity, C, (right axis) of CePd$_3$Mn$_{0.03}$
plotted as a function of temperature.

\end{description}

\end{document}